# Intensity asymmetries in the dusk sector of the poleward auroral oval due to IMF $B_x$

J. P. Reistad[1], N. Østgaard[1], K. M. Laundal[1,2], S. Haaland[1,3], P. Tenfjord[1], K. Snekvik[1], K. Oksavik[1,4], and S. E. Milan[1,5]

[1]Birkeland Centre for Space Science, Department of Physics and Technology, University of Bergen, Bergen, Norway, [2]Teknova AS, Kristiansand, Norway, [3]Max-Planck Institute for Solar System Research, Göttingen, Germany, [4]Department of Geophysics, University Centre in Svalbard, Longyearbyen, Norway, [5]Department of Physics and Astronomy, University of Leicester, Leicester, UK

**Abstract** In the exploration of global-scale features of the Earth's aurora, little attention has been given to the radial component of the Interplanetary Magnetic Field (IMF). This study investigates the global auroral response in both hemispheres when the IMF is southward and lies in the *xz* plane. We present a statistical study of the average auroral response in the 12–24 magnetic local time (MLT) sector to an *x* component in the IMF. Maps of auroral intensity in both hemispheres for two IMF $B_x$ dominated conditions (± IMF $B_x$) are shown during periods of negative IMF $B_z$, small IMF $B_y$, and local winter. This is obtained by using global imaging from the Wideband Imaging Camera on the IMAGE satellite. The analysis indicates a significant asymmetry between the two IMF $B_x$ dominated conditions in both hemispheres. In the Northern Hemisphere the aurora is brighter in the 15–19 MLT region during negative IMF $B_x$. In the Southern Hemisphere the aurora is brighter in the 16–20 MLT sector during positive IMF $B_x$. We interpret the results in the context of a more efficient solar wind dynamo in one hemisphere. Both the intensity asymmetry and its location are consistent with this idea. This has earlier been suggested from case studies of simultaneous observations of the aurora in both hemispheres, but hitherto never been observed to have a general impact on global auroral brightness in both hemispheres from a statistical study. The observed asymmetries between the two IMF $B_x$ cases are not large; however, the difference is significant with a 95% confidence level. As the solar wind conditions examined in the study are rather common (37% of the time) the accumulative effect of this small influence may be important for the total energy budget.

## 1. Introduction

To study how the Earth is coupled to space, information of how the two hemispheres respond differently to external forcing is of great interest. Simultaneous imaging from both hemispheres has been used to identify and investigate possible mechanisms responsible for the observed asymmetries of the global aurora [e.g., *Stenbaek-Nielsen and Davis*, 1972; *Craven et al.*, 1991; *Stenbaek-Nielsen and Otto*, 1997; *Sato et al.*, 1998; *Østgaard et al.*, 2003; *Fillingim et al.*, 2005; *Laundal and Østgaard*, 2009; *Østgaard et al.*, 2011; *Reistad et al.*, 2013].

Global imaging of the auroral oval represents a unique opportunity to look at footprints of a large region of the magnetosphere simultaneously. Although this has been possible for a long time on a sporadic basis from the mid-1980s, little is known about how the *x* component of the interplanetary magnetic field (IMF) can affect the aurora, especially in relation to its response in both hemispheres. Statistical studies of global auroral UV brightness dependence on solar wind (SW) and IMF in the Northern Hemisphere indicate that negative IMF $B_x$ on average produces brighter aurora than positive IMF $B_x$ during negative IMF $B_z$ conditions [*Liou et al.*, 1998; *Shue et al.*, 2002; *Baker et al.*, 2003].

Largely based on global auroral imaging, *Østgaard and Laundal* [2012] proposed that nonconjugate aurora, which is aurora only appearing in one end of the magnetic field line or is significantly brighter in one hemisphere, could be caused by three different generator mechanisms: the solar wind (SW) dynamo (related to the interplanetary magnetic field (IMF) $B_x$), effect of IMF $B_y$ penetration, and ionospheric conductivity. The recent study by *Reistad et al.* [2013] indicated that those three mechanisms indeed seem to play an important role in controlling the occurrence and location of nonconjugate aurorae.

In this paper we will go one step further and investigate if there is a significant auroral intensity difference in the two hemispheres in the dusk sector due to IMF $B_x$. As will be outlined in the paper, we apply certain selection criteria in order to avoid effects of other possible mechanisms that can produce asymmetric aurora. We focus on the dusk sector and especially the poleward part of the oval because this is the region where we expect the effects of IMF $B_x$ to be observable in our data. In this region we have typically upward field-aligned currents (Region 1) and precipitating electrons. As suggested earlier [*Cowley*, 1981; *Laundal and Østgaard*, 2009; *Reistad et al.*, 2013], hemispheric intensity asymmetries in the aurora in this region could be a signature of asymmetric Region 1 currents in the two hemispheres. In the discussion section we will look at our results in the light of this related work.

## 2. Data and Method
### 2.1. Global Imaging Data

To investigate how IMF $B_x$ affects the auroral intensity and distribution in both hemispheres, we have used the Far Ultraviolet Wideband Imaging Camera (WIC) [*Mende et al.*, 2000] on board the IMAGE (Imager for Magnetopause-to-Aurora Global Exploration) satellite [*Burch*, 2000]. IMAGE was launched in 2000 into an elliptic orbit with an apogee of 7 $R_E$ precessing over Northern Hemisphere at a rate of about 50° per year. From early 2004, the apsidal precession of the orbit allowed imaging of the Southern Hemisphere with almost the same coverage as it initially had after launch in March 2000. Thus, the IMAGE WIC data set represents a unique opportunity to study the average global response of the aurora statistically in both hemispheres with the same instrument.

The WIC instrument is sensitive to the Lyman-Birge-Hopfield band and a few N lines of the aurora in the UV range (140–190 nm) [*Mende et al.*, 2000]. In general, both precipitating electrons and protons produce these emissions and their relative influence on the resulting brightness depends primarily on their energy fluxes [*Frey et al.*, 2003]. Usually, the precipitating proton energy flux is too small to influence the WIC signal, but in some cases it can account for a significant portion of the signal [*Frey et al.*, 2001; *Donovan et al.*, 2012]. However, in the poleward part of the dusk sector oval this should not be a problem as we focus on the upward Region 1 current with limited downward proton fluxes. In the discussion we will use the intensity counts as measured by the WIC camera as a proxy for upward field-aligned current. This will be further explained in section 4.

### 2.2. Image Selection

We have used 1 min IMF and SW data from NASA's Space Physics Data Facility, http://omniweb.gsfc.nasa.gov [e.g., *King and Papitashvili*, 2005] represented in the Geocentric Solar Magnetospheric coordinate system. These data represent the conditions at the nose of the Earth's bow shock. To account for the additional propagation time needed for the IMF to affect the magnetosphere, we further shift the IMF data to $x = -10\ R_E$ using the present SW velocity. This is consistent with what has been done in earlier event studies, e.g., *Reistad et al.* [2013] and also what is suggested by MHD models regarding Region 1 current generation at the magnetopause [*Siscoe et al.*, 2000].

The event selection criteria are crucial when investigating the IMF $B_x$ effects. We want to exclude contributions from other possible mechanisms [*Østgaard and Laundal*, 2012] as far as statistical sample number allows for. We start by identifying intervals during the IMAGE mission where strict criteria on IMF and seasonal variations are met. Only global imaging data that fulfils the following criteria are used: (1) |IMF $B_x$| > 2 nT, (2) |IMF $B_y$| < 2 nT, (3) IMF $B_z$ < 0 nT, (4) 10° < |Dipole tilt| < 30°, (5) > 10 min between observations, and (6) criteria must be satisfied for more than 10 min. In the identified time intervals, a global image is chosen toward the end of the interval but not less than 5 min before the end of the interval. This is to ensure that we select images that have been exposed to the favorable IMF and tilt criteria as long as possible. Using these criteria we get two sets of images in each hemisphere which will be compared: One set for IMF $B_x < -2$ nT, and one set for IMF $B_x > 2$ nT.

We want to exclude, as good as possible, other mechanisms that can produce asymmetric aurora to avoid that the IMF $B_x$ signatures drown in other stronger signals. The selection process is therefore crucial and makes this study different from other studies that have performed statistical analysis on a much wider spectrum of geomagnetic conditions [e.g., *Shue et al.*, 2001, 2002; *Baker et al.*, 2003].

The negative IMF $B_z$ condition is chosen to include only the intervals where magnetic flux is opened on the dayside and convected across the polar cap. At the same time we want to include only times when there is a significant $x$ component and a small $y$ component in the IMF to minimize signatures of IMF $B_y$ effects. Due to seasonal effects and the offset of the geomagnetic pole from the rotational axis, the two polar ionospheres are exposed differently to solar radiation in magnetic coordinates. The difference is to the first order quantified by the dipole tilt angle. Due to a large contribution of dayglow-induced emissions in the WIC camera from regions in direct sunlight, we only use images from the local winter hemisphere. In the Northern Hemisphere we use images having dipole tilt angles from $-30°$ to $-10°$, and in the Southern Hemisphere we use only images having dipole tilt angle from $10°$ to $30°$. The use of local winter periods allows us to more accurately examine differences in a large region of the auroral oval extending to $\sim$ 15 MLT on the duskside in order to cover more of the Region 1 current region in the dusk sector. As mentioned in section 1, seasonal differences may lead to the occurrence of asymmetric aurora. However, these effects should not depend upon IMF $B_x$. When applying the same tilt criteria in both hemispheres such effects should not affect our results when having a sufficient number of samples.

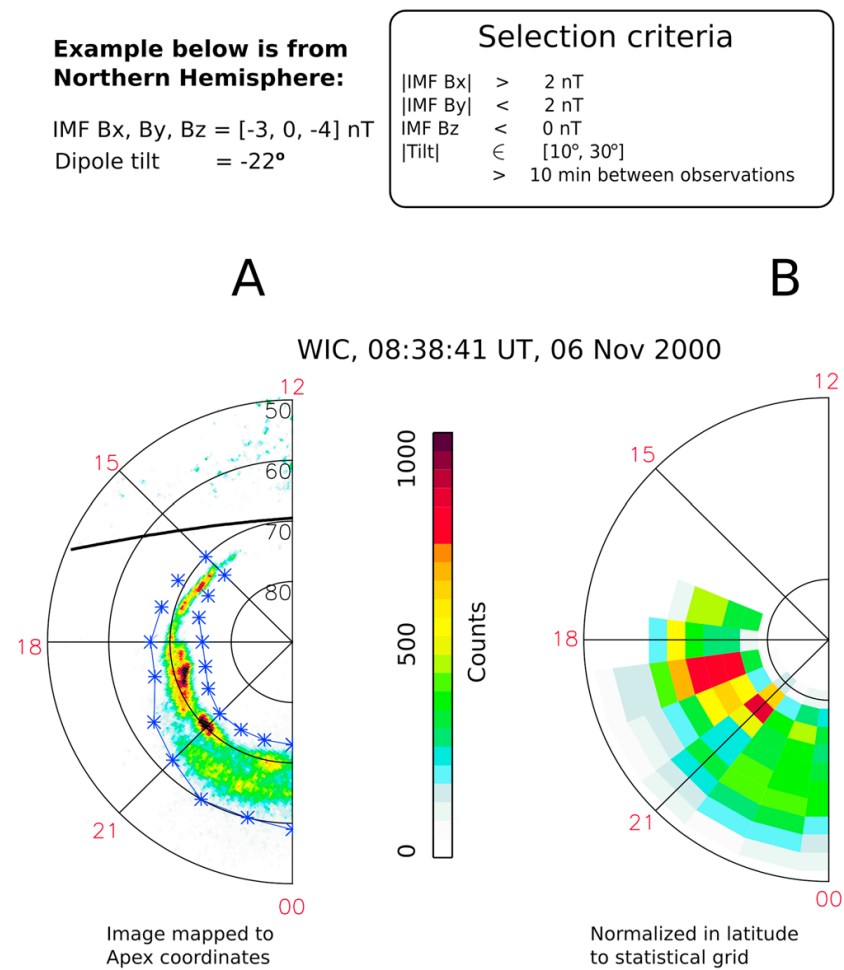

**Figure 1.** Selection criteria and overview of the method for image processing. (a) Example from the Northern Hemisphere after dayglow subtraction satisfying the selection criteria. Blue lines indicate the detected oval boundaries with asterisks located at the center of the 1 h MLT bin used in the statistics. (b) MLT sectors where all criteria are met get normalized to a common 10 bin latitudinal extent. The values seen in Figure 1b enter the statistical analysis.

The WIC images have a 2 min cadence, and consecutive images are therefore correlated. In the statistical analysis we require observations to be more than 10 min apart to be considered not correlated. This reduces the number of observations but gives a set of images where each event is weighted more equally.

### 2.3. Image Processing

Although we require the images to be taken during local winter as quantified by criterion (4) listed above, some regions of the auroral oval are still directly exposed to sunlight leaving dayglow-induced emissions in the WIC images. The dayglow-induced emissions, and a varying background, are therefore subtracted from each image separately. This is done by constructing a model of the dayglow emissions from pixels not influenced by aurora based on their solar zenith angle and satellite zenith angle in the mapped image. The modeled pixel intensity is then subtracted from all pixels leaving only auroral emissions in the image. This introduces an uncertainty in the regions exposed to sunlight, especially regions far into the dayside, typically the 12–14 MLT sector in our data set. Strict criteria in the dayglow removal technique and manual inspection of the performance for each image result in less data from these regions. For the chosen range of dipole tilt angles, the oval will on average be in darkness from 15 MLT and tailward on the duskside. Therefore, the analysis will focus on the 15–24 MLT region. An example of an image from the Northern Hemisphere satisfying the selection criteria after dayglow removal is shown in Figure 1a. The black line oriented in a dusk-dawn direction shows the location of the terminator. Since we investigate the possible IMF $B_x$ influence on the duskside auroral intensity, we only show this sector as it corresponds to the upward Region 1 current and electron precipitation into the ionosphere.

As we want to study the intensity within the oval, the images are transformed into a frame following the oval. The advantage is that the statistics will be less affected by the location of the oval from event to event and minor errors in the satellite pointing direction. Previous statistical studies focusing on average auroral intensities or energies [e.g., *Shue et al.*, 2001, 2002; *Newell et al.*, 1996] superposed their data on a fixed MLT/MLAT grid. By using a normalized latitudinal width of the oval, signatures at the poleward/equatorward boundary and within the oval can be more easily identified. To do so, we determine the poleward and equatorward boundaries of the oval as shown in Figure 1a. The blue lines with asterisks indicate the oval boundary and the 1 h MLT resolution that we use in the statistical grid. The boundaries are determined by a threshold value depending on the mean and the spread of the counts in a 1 h by 1° MLT/MLAT grid. When the boundaries have been determined, the auroral signatures in MLT sectors with valid boundaries are normalized into a common latitudinal extent of 10 entries. To be considered as valid boundaries we require the poleward boundary to be between 62° and 81° latitude, and the equatorward boundary to be between 52° and 72°. In addition, we require that neighboring MLT sector boundaries within one image should not differ by more than 4° and that the total width of the oval should not exceed 15° MLAT. For the event shown in Figure 1a, all the criteria are satisfied and the corresponding aurora normalized in MLT sectors are rebinned in 10 latitudinal bins as shown in Figure 1b. As a result, each image can contribute with up to 13 MLT slices to the statistical analysis. When discussing the results, the number of events is referring to the number of such slices.

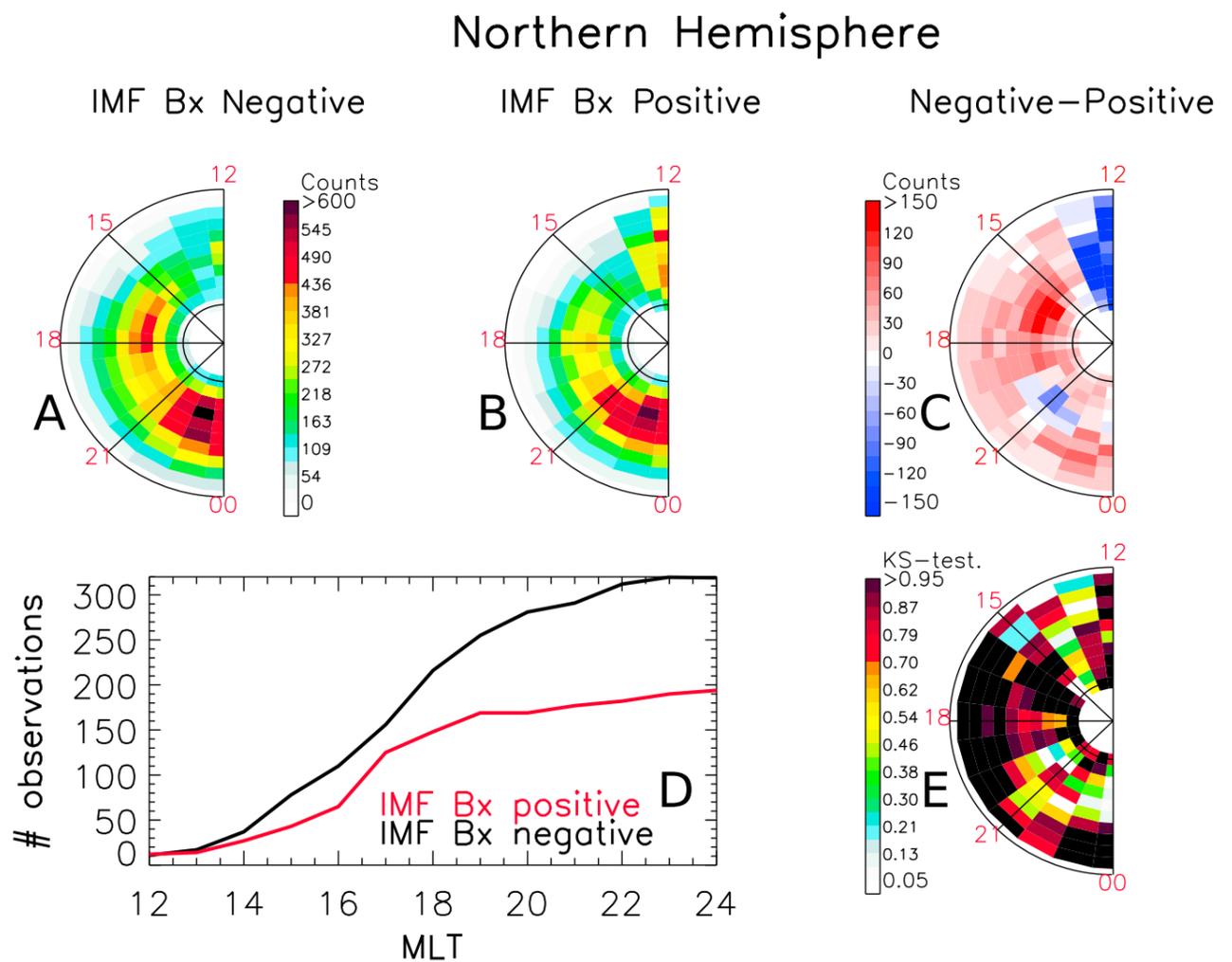

**Figure 2.** Results of the statistical analysis in the Northern Hemisphere. (a) Maps of median auroral intensity for the IMF $B_x$ negative case. (b) Corresponding map for the positive IMF $B_x$ case. (c) The intensity difference between the two IMF $B_x$ cases. (d) Number of events across the domain. (e) The significance level of the test that the two distributions do not originate from the same distributions, using the Kolmogorov-Smirnov test. The expected increased brightness for negative IMF $B_x$ can be seen from 16 to 19 MLT.

### 2.4. Statistical Analysis

When showing maps of auroral intensities for the two IMF $B_x$ sets in each hemisphere, we use the median of the distribution. This will reduce the weighting of extreme events. A similar result is obtained when using the mean (not shown here). To test the hypothesis that the distribution of data for a given MLT value and latitudinal bin is different for the two different IMF $B_x$ cases in the same hemisphere, we perform a Kolmogorov-Smirnov test [*Press et al.*, 1992]. Based on this test, we identify in which regions of the maps the hypothesis is found to be true (data are not drawn from the same distribution), given a significance level. To show how the significance level varies in the grid, we present maps of the significance level.

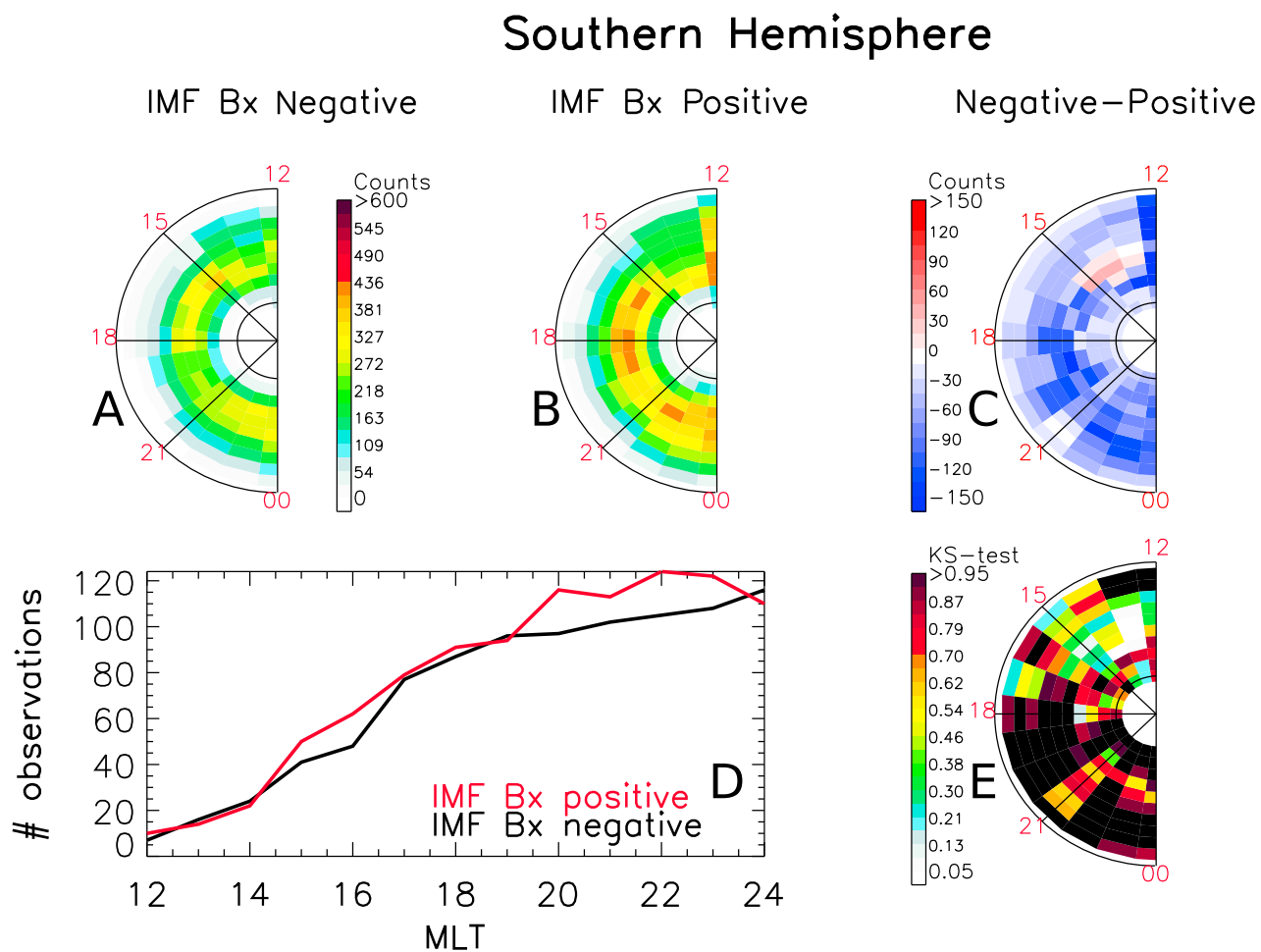

**Figure 3.** Results of the statistical analysis in the Southern Hemisphere. (a) Maps of median auroral intensity for the IMF $B_x$ negative case. (b) Corresponding map for the positive IMF $B_x$ case. (c) The intensity difference between the two IMF $B_x$ cases. (d) Number of events across the domain. (e) The significance level of the test that the two distributions do not originate from the same distributions, using the Kolmogorov-Smirnov test. The expected increased brightness for positive IMF $B_x$ can be seen from 16 to 20 MLT.

## 3. Results

Figures 2 and 3 show the results from the statistical analysis in the same format for the Northern and Southern Hemispheres, respectively. Figure 2a shows the dusk sector (12–24 MLT) median auroral intensity for the negative IMF $B_x$ case. The corresponding plot for the positive IMF $B_x$ case is shown in Figure 2b and their difference in Figure 2c. The map in Figure 2e shows the probability that the two distributions in every cell in Figures 2a and 2b are not drawn from the same distribution by using the Kolmogorov-Smirnov test. The color scale is chosen to show how the significance varies in regions outside the 95% (5%) confidence interval. In the black regions the hypothesis is true (there is an asymmetry), and in white regions it is falsified (there is not an asymmetry). Figure 2d shows how many events that were used in the statistics for the two cases in different MLT sectors. A line plot is here provided as there is an equal number of events in each latitudinal bin for a given MLT sector due to the oval boundary and normalization method used.

For the Northern Hemisphere results in Figure 2 the 15–19 MLT range for the IMF $B_x$ negative case is more intense than the IMF $B_x$ positive case. The signature is also in the poleward half of the normalized oval and is significant on a 95% confidence level in the 15–17 MLT sector. The contributions from substorm activity are clearly seen with a peak around 23 MLT, as expected. However, there are no significant asymmetries between the two IMF $B_x$ cases in this region.

The results for the corresponding situation in the Southern Hemisphere during the same local winter conditions are seen in Figure 3 in the same format as for the Northern Hemisphere. Now the positive IMF $B_x$ case in Figure 3b shows the most intense aurora in the 16–20 MLT sector. Taking the data spread and number of observations into account, the Kolmogorov-Smirnov test indicates that the 17–20 MLT sector is more than 95% likely to be drawn from different distributions, meaning that there is a significant IMF $B_x$ influence in this region also in the Southern Hemisphere.

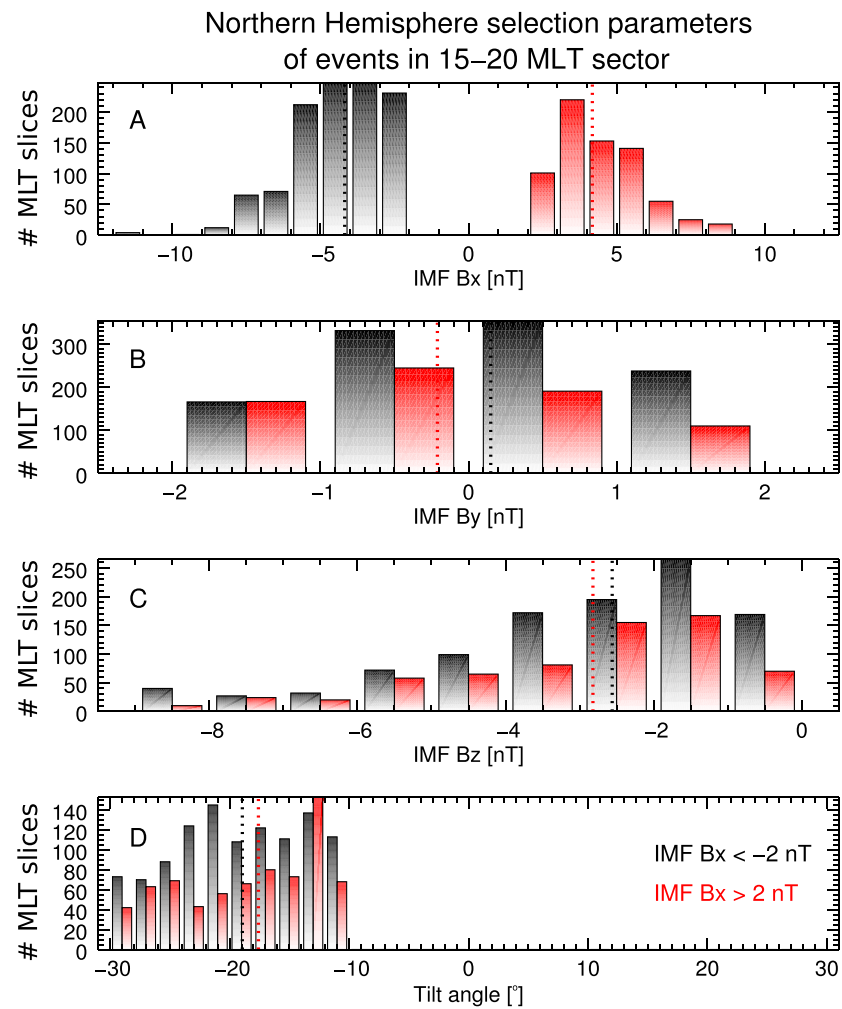

**Figure 4.** Distribution of IMF and dipole tilt angles for the events in the statistical study for the Northern Hemisphere. Black bars represent the negative IMF $B_x$ case, and red bars the positive IMF $B_x$ case. Number of MLT slices refers to the total number of events in the 15–20 MLT sector for the given parameter interval and IMF $B_x$ case. Vertical dashed lines represent the median of the distribution. (a) IMF $B_x$ distribution, (b) IMF $B_y$ distribution, (c) IMF $B_z$ distribution, and (d) dipole tilt angle distribution.

## 4. Discussion

As seen in Figures 2e and 3e, there is a region 15–17 MLT in Northern Hemisphere and 17–20 MLT in Southern Hemisphere, where the brightness for the two IMF $B_x$ cases differs significantly. In the Northern Hemisphere the difference in the 15–19 MLT region is about 170 camera counts or ~0.3 kR in the WIC passband. In the Southern Hemisphere the difference is ~150 camera counts for the 16–20 MLT sector, well above the Poisson noise in the images with a standard deviation of typically 30 counts. Typical count rates can be seen in Figure 1a. This difference is a relatively small modification on the general auroral brightness, as also expected as |IMF $B_x$| > 2 nT is not a strong criterion. However, it is significant in our data set which is chosen to include the periods when IMF $B_x$ asymmetries are believed to be most prominent. The selection criteria used to avoid IMF $B_y$ influences on our result will also rule out periods of strong driving since such periods often have an |IMF $B_y$| > 2 nT. We also tried a different criteria, requiring IMF $B_x$ to be twice the magnitude of IMF $B_y$ and no limitation on their magnitude. The results were similar but slightly less significant indicating contamination from IMF $B_y$ effects.

When comparing the two IMF $_B x$ cases it is important that both distributions on average represent the same geomagnetic conditions. In particular the IMF $B_z$ is important when looking at the median auroral response as this will affect the overall driving of the system. In Figures 4 and 5 we present histograms for the IMF components and tilt angles concurrent with each latitudinal slice in the 15–20 MLT sector that enter into the statistics in Figures 2 and 3. The negative IMF $B_x$ case is shown in black and positive IMF $B_x$ in red. The median of the distributions is shown with vertical dashed lines in respective colors. In Figure 4c we see that the shape of the Northern Hemisphere distributions of IMF $B_z$ is similar for both IMF $B_x$ cases, and their median value differs by 0.3 nT. This is a small difference. We also notice that the IMF $B_x/B_y$ antisymmetry is almost completely avoided due to the selection criteria. The median of the two distributions lies within ±0.2 nT as seen in Figure 4b. Hence, we conclude that the Northern Hemisphere data sets are not biased with respect to SW driving. The difference of the median dipole tilt angle is also low, about 1°. The main difference between the two cases is that there are fewer IMF $B_x$ positive events.

In the Southern Hemisphere we see from Figures 5a and 5b that the IMF $B_x/B_y$ antisymmetry only results in a small shift of around 0.4 nT from 0 of the median for IMF $B_y$. Looking at the IMF $B_z$ distributions in Figure 5c, the difference of the median is now slightly larger for the two IMF $B_x$ cases compared to the Northern Hemisphere (Figure 4c). The difference is here 0.5 nT in the direction of the observed asymmetry, making the Southern Hemisphere results slightly biased from this point of view. We have investigated this possible bias by looking at how the count rate depend upon IMF $B_z$. From every MLT slice in the 15–20 MLT sector, a scatterplot of the mean count versus IMF $B_z$ was made. From this scatter, the median count value of points within 1 nT wide bins of IMF $B_z$ gives a line that relates how strongly the different events respond. Such lines are shown for the Northern Hemisphere in Figure 6 for both IMF $B_x$ cases (negative IMF $B_x$ in black and positive IMF $B_x$ in red). We can see that the negative IMF $B_x$ case has a stronger response for all IMF $B_z$, consistent with our results in Figure 2. The histogram from Figure 4c is also plotted in the same figure, and we can see how the 95% confidence level of the median [*Chen and Ratra*, 2011] increases with decreasing number of MLT slices. Looking at the corresponding plot for the Southern Hemisphere data, shown in Figure 7, one can see that the general IMF $B_x$ difference we presented in Figure 3 is evident in the IMF $B_z \in [-4, 0]$ nT range. Looking at the IMF $B_z$ histogram in the same plot we see that the majority of the events occur for IMF $B_z > -4$ nT. It is evident from Figure 7 that the few events having large negative IMF $B_z$ do not affect the results in a way that will give larger intensities for the positive IMF $B_x$ case. We hence conclude that the observed IMF $B_x$ difference in the Southern Hemisphere cannot be explained by a bias in selection regarding IMF $B_z$.

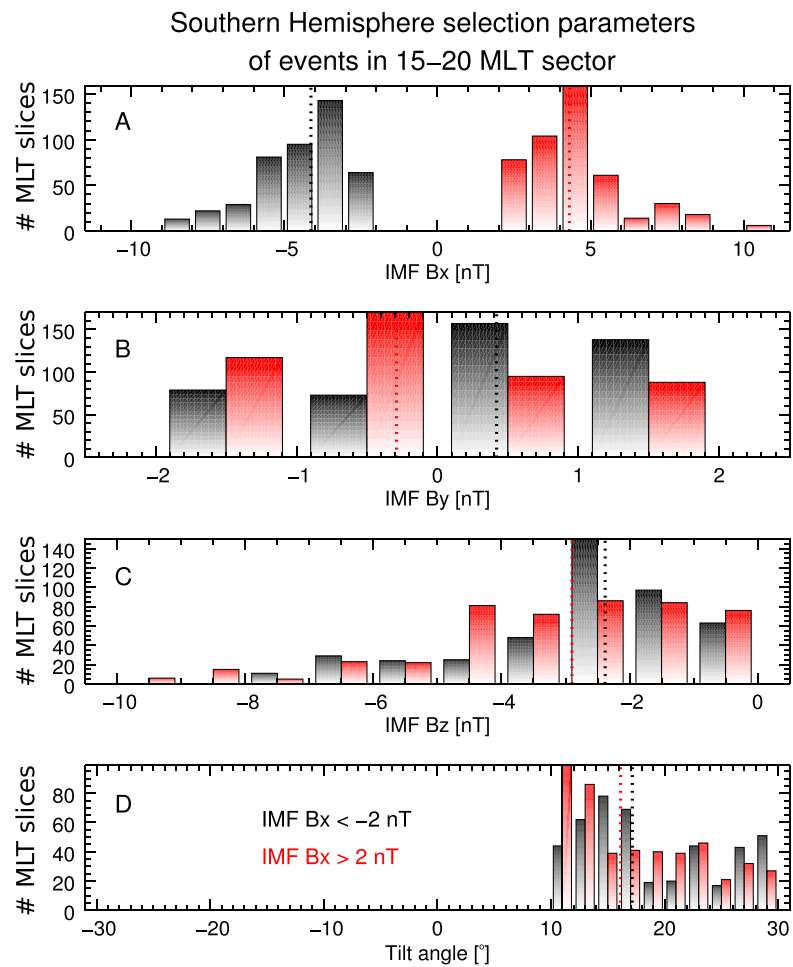

**Figure 5.** Distribution of IMF and dipole tilt angles for the events in the statistical study for the Southern Hemisphere. The figure is in the same format as Figure 4.

As mentioned in section 2, the measured WIC signal primarily originates from precipitating electrons. In the ionosphere, the upward Region 1 current is associated with electron precipitation [e.g., *Paschmann et al.*, 2002; *Mende et al.*, 2003a, 2003b; *Dubyagin et al.*, 2003]. Hence, the dusk sector Region 1 current is of interest. However, WIC is only sensitive to the accelerated part of the electron spectra, typically 0.5–5 keV. Within this energy range the WIC counts are nearly proportional to the electron energy flux [*Frey et al.*, 2003]. In general, the current could be carried by lower energy electrons leaving little to no signatures in auroral UV brightness. Within this sector and outside the regions associated with magnetic reconnection the precipitation is characterized by so-called "inverted Vs" [*Newell et al.*, 1996; *Chaston et al.*, 2007]. The inverted-V type of precipitation is the most dominant in the local winter hemisphere dusk region [*Newell et al.*, 2009] and is believed to carry the majority of the upward field-aligned current. Another characteristic electron spectrum is the Alfvén wave accelerated spectrum, characterized by a broad energy distribution and associated bright aurora [*Chaston et al.*, 2003]. Such precipitation is found to occur typically in regions associated with magnetic reconnection [*Chaston et al.*, 2007] and substorm aurora [*Mende et al.*, 2003a; *Newell et al.*, 2009]; and hence, its signature is expected to have a fundamentally different origin. Using particle and field measurements from low-Earth orbiting satellites, [*Ohtani et al.*, 2009] have shown that the dusk sector Region 1 current density is proportional to the precipitating electron energy flux. They found the mean electron energy during these crossings in darkness to be typically between 1.5 and 2 keV which is the electron energy range the WIC instrument is most sensitive to. Hence, event observations of simultaneous global aurora [*Laundal and Østgaard*, 2009; *Reistad et al.*, 2013] showing increased ionospheric UV brightness in the 18–24 MLT region are consistent with the Region 1 current system having different intensity in the two hemispheres. Based on this we will interpret our statistical auroral response in the 15–20 MLT sector during local winter conditions as a proxy of the average Region 1 currents.

The question regarding where and how the Region 1 currents are generated at the magnetopause and coupled to the ionosphere has been investigated extensively [e.g., *Siscoe et al.*, 1991, 2000; *Song and Lysak*, 2001; *Guo et al.*, 2008; *Lopez et al.*, 2011]. MHD modeling studies such as *Siscoe et al.* [2000] and *Guo et al.* [2008] suggest that the Region 1 current at most local times tends to be driven across the magnetopause. From these arguments we would expect to see a possible influence on the aurora where the Region 1 current is flowing out of the ionosphere, namely in a large portion of the 12–24 MLT sector. In Figures 2 and 3 we have chosen to include periods with substorms. These periods are associated with IMF $B_z$ negative and therefore strong Region 1 currents. Leaving out substorm periods using the list of substorms also identified by the WIC camera [*Frey et al.*, 2004] and only use data more than 90 min after substorm onset, we get the same trends in our results but not the same significance levels. Although substorm intervals represent a contamination of our data we will argue that the Region 1 currents have a different origin and footprint than substorms. As long as we identify the difference within the 15–20 MLT sector, we can interpret our results to reflect the possible IMF $B_x$ influences on the directly driven Region 1 currents.

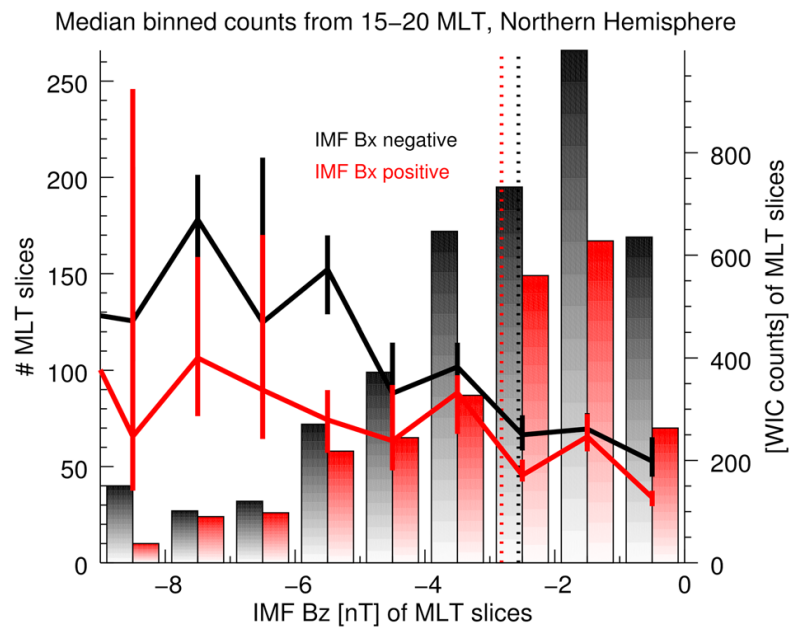

**Figure 6.** Line plots of WIC response for different IMF $B_z$ values in the data set derived from median binned values of average counts for each event in the 15–20 MLT sector in the Northern Hemisphere. Histograms of IMF $B_z$ are also shown for comparison. Black and red lines and bars represent the negative and positive IMF $B_x$ cases, respectively. Axis to the left corresponds to the histogram, and axis to the right corresponds to the lines.

In the following we will interpret our findings in the context of how the SW dynamo can have different efficiency in the two hemispheres due to an IMF $B_x$ component. By asymmetric SW dynamo efficiency we mean that the energy transfer from the SW to the magnetosphere can be different in the two hemispheres.

As first suggested by *Cowley* [1981] and later supported by others [e.g., *Laundal and Østgaard*, 2009; *Østgaard and Laundal*, 2012; *Reistad et al.*, 2013], an $x$ component of the IMF could, during negative IMF $B_z$ conditions, lead to different magnetic tension forces acting on the open magnetic field lines being draped down tail, possibly affecting the energy transfer at the magnetopause differently in the two hemispheres. This effect is illustrated in Figures 8a and 8b for both negative and positive IMF $B_x$, respectively, during southward IMF $B_z$ and small IMF $B_y$. Figures 8a and 8b are a remake of the original figure from *Cowley* [1981].

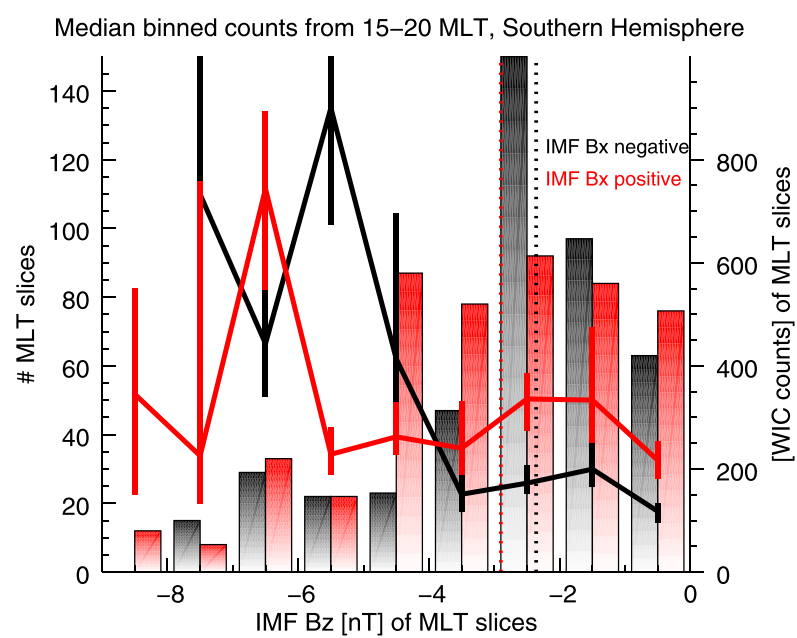

**Figure 7.** Line plots of WIC response for different IMF $B_z$ values in the data set derived from median binned values of average counts for each event in the 15–50 MLT sector in the Southern Hemisphere. Histograms of IMF $B_z$ are also shown for comparison. Black and red lines and bars represent the negative and positive IMF $B_x$ cases, respectively. Axis to the left corresponds to the histogram, and axis to the right corresponds to the lines.

Here the magnitude of the magnetic tension force on open field lines is illustrated with wide black arrows where the field lines have a small curvature radius, and smaller black arrows in the opposite hemisphere experiencing a smaller tension force due to the greater radius of curvature. The SW electric field, due to the SW velocity and IMF, during the conditions illustrated in Figure 8, points out of the paper as indicated by the red arrows. In the high-latitude region tailward of the terminator the electric field and $\mathbf{j}_\perp$ point in opposite directions, hence the name SW dynamo. When neglecting particle pressure gradients in the magnetohydrodynamic (MHD) momentum equation for a quasi-neutral plasma, and rewriting the $\mathbf{j} \times \mathbf{B}$ force, we are left with one force term that depends on the curvature of the magnetic field—the magnetic tension force—and one related to the gradient in

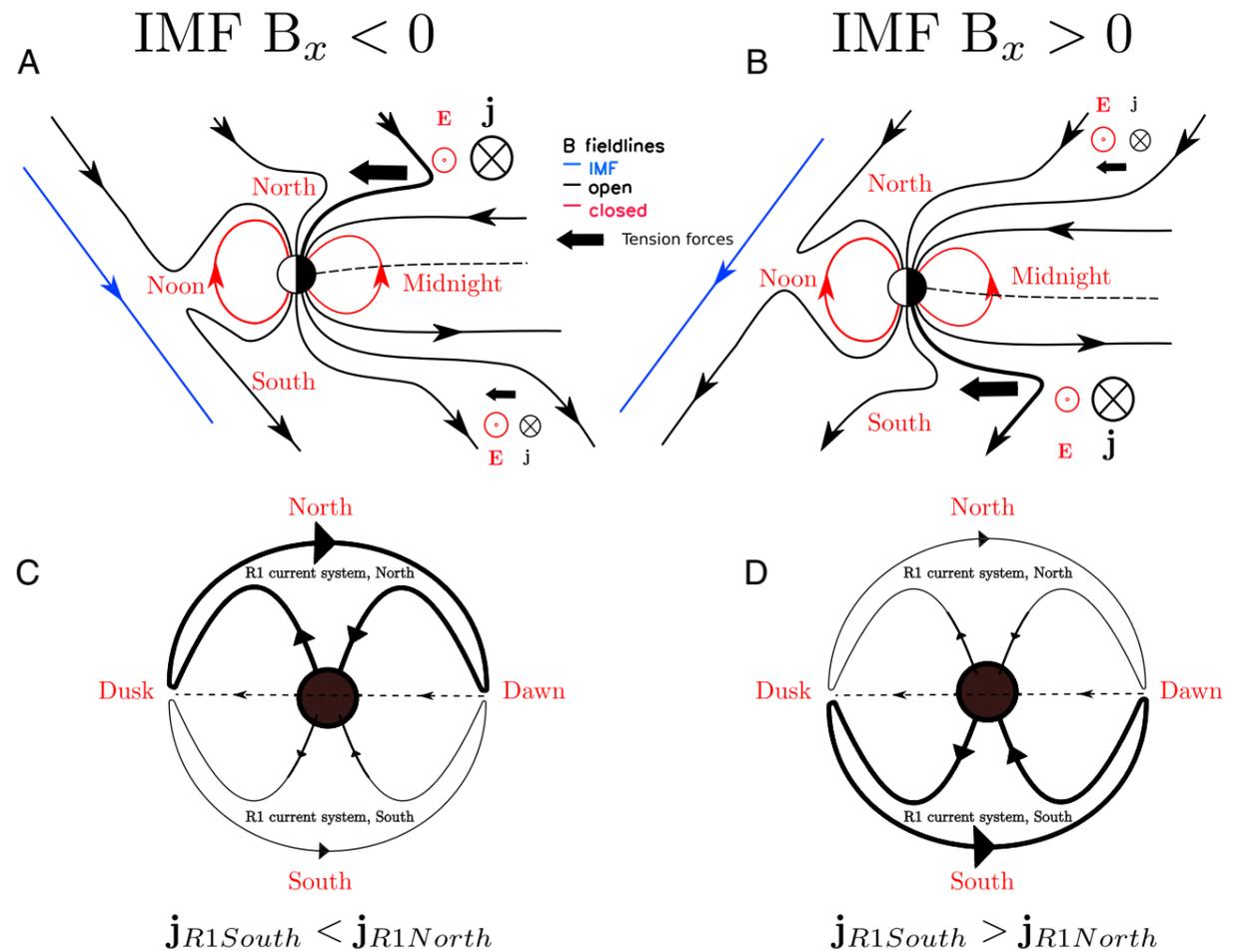

**Figure 8.** (a) The presence of a negative *x* component in the IMF during negative IMF $B_z$ resulting in different magnetic tension forces (illustrated with large black arrows) on open field lines possibly affecting the energy transfer at the magnetopause different in the two hemispheres. Lines in blue, black, and red show the interplanetary, open, and closed field lines, respectively. At high latitudes, tailward of the terminator, the tension force will decelerate plasma and affect the magnetopause current density. This is indicated with the black arrow going into the figure in this region. This figure is adopted from *Cowley* [1981]. (b) Corresponding figure during positive IMF $B_x$. (c) Associated changes in current density for negative IMF $B_x$ of the Region 1 current system that closes across the high-latitude magnetopause [*Siscoe et al.*, 1991; *Siscoe et al.*, 2000; *Guo et al.*, 2008]. Here the Earth is viewed from the tail. (d) Corresponding figure during positive IMF $B_x$.

magnetic pressure perpendicular to the magnetic field:

$$\rho_m \frac{d\mathbf{v}}{dt} = \mathbf{j} \times \mathbf{B} = \frac{B^2}{\mu_0 R_c}\hat{\mathbf{n}} - \nabla_\perp \left(\frac{B^2}{2\mu_0}\right), \qquad (1)$$

where $\rho_m$ is mass density, $R_c$ is the radius of curvature of the magnetic field, $\hat{\mathbf{n}}$ is the unit normal pointing toward the center of curvature, and $\frac{d\mathbf{v}}{dt}$ is the total derivative of the plasma velocity. Using vector identities and Ampère's law, equation (1) takes the form

$$\mathbf{j}_\perp = \frac{\rho_m \mathbf{B} \times \frac{d\mathbf{v}}{dt}}{B^2}. \qquad (2)$$

Applied to the SW-magnetosphere system, $\mathbf{j}_\perp$ represents the magnetopause current density normal to the magnetic field resulting from the combined effect of the two terms on the right side in equation (1). MHD modeling of magnetospheric current systems [e.g., *Siscoe et al.*, 2000; *Tanaka*, 2000; *Guo et al.*, 2008] indicates that the Region 1 currents enclosing the polar cap [*Iijima and Potemra*, 1978] indeed seem to be generated on the high-latitude magnetopause as perpendicular currents, as earlier suggested by e.g. *Siscoe et al.* [1991] in their Figure 1. This is conceptually indicated in Figures 8c and 8d where the Earth is viewed from the tail. Taking into account the possible change of magnetic tension force on the open field lines due to an IMF $B_x$ component, this could alter the magnitude of the first term on the right side in equation (1) and hence affect $\mathbf{j}$. For negative IMF $B_x$ (Figure 8a) open field lines on the nightside are conceptually believed to experience a stronger magnetic tension force in the Northern Hemisphere. This can eventually increase $\mathbf{j}$ in the Northern Hemisphere relative to the Southern Hemisphere according to equation (1). For positive IMF $B_x$ it will be the other way around as shown in Figure 8b by the black arrows going into the paper. For the suggested generation of Region 1 currents at the magnetopause [*Siscoe et al.*, 1991, 2000; *Tanaka*, 2000; *Guo et al.*, 2008], this asymmetric influence on **j** due to IMF $B_x$ could affect the Region 1 currents differently in the two hemispheres. This is illustrated for both IMF $B_x$ cases in Figures 8c and 8d by the different thickness of the lines in the two hemispheres.

The observations presented in Figure 2 are consistent with an enhanced SW dynamo in the Northern Hemisphere during IMF $B_x$ negative as illustrated in Figures 8a and 8c. The signature is also in the poleward half of the normalized oval, consistent with the Region 1 current location. The observations presented from the Southern Hemisphere in Figure 3 are also in agreement with this idea, suggesting an enhanced SW dynamo in the Southern Hemisphere during positive IMF $B_x$. To the best of our knowledge, this is the first statistical observational study indicating that IMF $B_x$ can modify the energy conversion between the SW and the magnetosphere differently in the two hemispheres in a general sense, and although the average effect is found to be small the accumulative effect may be important. The |IMF $B_x$| > 2 nT criterion is in fact met 73% of the time during IMF $B_z$ < 0 conditions in the period 2000–2005.

The asymmetric SW dynamo has been suggested as a possible explanation for asymmetric aurora in earlier case studies [*Laundal and Østgaard*, 2009; *Reistad et al.*, 2013] but has not been mentioned in earlier statistical studies of aurora with regard to IMF $B_x$ [*Liou et al.*, 1998; *Shue et al.*, 2002; *Baker et al.*, 2003]. However, the IMF $B_x$ asymmetries found in the Northern Hemisphere in these previous statistical studies are consistent with our results, being generally stronger aurora in the Northern Hemisphere for negative IMF $B_x$. These results are therefore also consistent with what we would expect from increased SW dynamo action in the Northern Hemisphere during negative IMF $B_x$.

We emphasize that the observations and interpretation presented here are only indirect evidence of the influence of an asymmetric SW dynamo on auroral brightness. A more direct investigation during carefully selected intervals may give more insight into the importance of the IMF $B_x$ on SW dynamo efficiency.

## 5. Conclusion

We have shown median auroral intensity maps from the dusk sector in both hemispheres during negative IMF $B_z$ and dominating IMF $B_x$ over $B_y$ conditions. We have found that in the Northern Hemisphere the aurora is brighter in the 15–19 MLT sector during IMF $B_x$ negative conditions. This asymmetry is most evident in the poleward half of the indicated oval. In the Southern Hemisphere we observe an opposite behavior where the aurora in the 16–20 MLT sector is brighter during IMF $B_x$ positive conditions. In both hemispheres the two auroral distributions are significantly different with a 95% significance level within most of the indicated regions. Our results are consistent with an increased SW dynamo efficiency in the Northern Hemisphere for negative IMF $B_x$ and in the Southern Hemisphere for positive IMF $B_x$. The possible modulation of SW dynamo efficiency between the hemispheres due to IMF $B_x$ has only been suggested in earlier case studies. This is the first statistical study of observations from both hemispheres indicating that this mechanism is likely to produce general differences in auroral brightness. Although the asymmetries are weak in a general sense, the accumulative effect may be important in the total energy budget as |IMF $B_x$| > 2 nT is a common condition in the SW.

**Acknowledgments**
We thank S.B. Mende and the IMAGE FUV team for the use of IMAGE FUV data. We acknowledge the use of NASA/GSFC's Space Physics Data Facility for OMNI data. This study was supported by the Research Council of Norway under contract 223252/F50 and 212014/F50.

Yuming Wang thanks Jianpeng Guo and another reviewer for their assistance in evaluating this paper.

### References

Baker, J. B., A. J. Ridley, V. O. Papitashvili, and C. R. Clauer (2003), The dependence of winter aurora on interplanetary parameters, *J. Geophys. Res.*, 108(A4), 8009, doi:10.1029/2002JA009352.
Burch, J. L. (2000), Image mission overview, *Space Sci. Rev.*, 91, 1–14.
Chaston, C. C., L. M. Peticolas, J. W. Bonnell, C. W. Carlson, R. E. Ergun, J. P. McFadden, and R. J. Strangeway (2003), Width and brightness of auroral arcs driven by inertial Alfvén waves, *J. Geophys. Res.*, 108(A2), 1091, doi:10.1029/2001JA007537.
Chaston, C. C., C. W. Carlson, J. P. McFadden, R. E. Ergun, and R. J. Strangeway (2007), How important are dispersive Alfvén waves for auroral particle acceleration?, *Geophys. Res. Lett.*, 34, L07101, doi:10.1029/2006GL029144.
Chen, G., and B. Ratra (2011), Median statistics and the hubble constant, *Astron. Soc. Pac.*, 123(907), 1127–1132.
Cowley, S. W. H. (1981), Asymmetry effects associated with the x-component of the IMF in a magnetically open magnetosphere, *Planet. Space Sci.*, 29(8), 809–818.
Craven, J. D., J. S. Murphree, L. A. Frank, and L. L. Cogger (1991), Simultaneous optical observations of transpolar arcs in the two polar caps, *Geophys. Res. Lett.*, 18(12), 2297–2300.
Donovan, E., E. Spanswick, J. Liang, J. Grant, B. Jackel, and M. Greffen (2012), Magnetospheric dynamics and the proton aurora, in *Auroral Phenomenology and Magnetospheric Processes: Earth and Other Planets, Geophys. Monogr. Ser.*, vol. 197, edited by A. Keiling et al., pp. 99–111, AGU, Washington, D. C., doi:10.1029/2011GM001190.
Dubyagin, S. V., V. A. Sergeev, C. W. Carlson, S. R. Marple, T. I. Pulkkinen, and A. G. Yahnin (2003), Evidence of near-Earth breakup location, *Geophys. Res. Lett.*, 30(6), 1282, doi:10.1029/2002GL016569.

Fillingim, M. O., G. K. Parks, H. U. Frey, T. J. Immel, and S. B. Mende (2005), Hemispheric asymmetry of the afternoon electron aurora, *Geophys. Res. Lett.*, *32*, L03113, doi:10.1029/2004GL021635.

Frey, H. U., S. B. Mende, C. W. Carlson, J.-C. Gèrard, B. Hubert, J. Spann, R. Gladstone, and T. J. Immel (2001), The electron and proton aurora as seen by IMAGE-FUV and FAST, *Geophys. Res. Lett.*, *28*(6), 1135–1138.

Frey, H. U., S. B. Mende, T. J. Immel, J.-C. Gèrard, B. Hubert, S. Habraken, J. Spann, G. R. Gladstone, D. V. Bisikalo, and V. I. Shematovich (2003), Summary of quantitative interpretation of IMAGE far ultraviolet auroral data, *Space Sci. Rev.*, *109*, 255–283.

Frey, H. U., S. B. Mende, V. Angelopoulos, and E. F. Donovan (2004), Substorm onset observations by IMAGE-FUV, *J. Geophys. Res.*, *109*, A10304, doi:10.1029/2004JA010607.

Guo, X. C., C. Wang, Y. Q. Hu, and J. R. Kan (2008), Bow shock contributions to region 1 field-aligned current: A new result from global MHD simulations, *Geophys. Res. Lett.*, *35*, L03108, doi:10.1029/2007GL032713.

Iijima, T., and T. A. Potemra (1978), Large-scale characteristics of field-aligned currents associated with substorms, *J. Geophys. Res.*, *83*(A2), 599–615, doi:10.1029/JA083iA02p00599.

King, J. H., and N. E. Papitashvili (2005), Solar wind spatial scales in and comparisons of hourly Wind and ACE plasma and magnetic field data, *J. Geophys. Res.*, *110*, A02104, doi:10.1029/2004JA010649.

Laundal, K. M., and N. Østgaard (2009), Asymmetric auroral intensities in the Earth's Northern and Southern Hemispheres, *Nature*, *460*(7254), 491–493, doi:10.1038/nature08154.

Liou, K., P. T. Newell, C. I. Meng, M. Brittnacher, and G. Parks (1998), Characteristics of the solar wind controlled auroral emissions, *J. Geophys. Res.*, *103*(A8), 17,543–17,557.

Lopez, R. E., V. G. Merkin, and J. G. Lyon (2011), The role of the bow shock in solar wind-magnetosphere coupling, *Ann. Geophys.*, *29*, 1129–1135, doi:10.5194/angeo-29-1129-2011.

Mende, S. B., et al. (2000), Far ultraviolet imaging from the IMAGE spacecraft. 2. Wideband FUV imaging, *Space Sci. Rev.*, *91*, 271–285.

Mende, S. B., C. W. Carlson, H. U. Frey, T. J. Immel, and J.-C. Gèrard (2003a), IMAGE FUV and in situ FAST particle observations of substorm aurorae, *J. Geophys. Res.*, *108*(A4), 8010, doi:10.1029/2002JA009413.

Mende, S. B., C. W. Carlson, H. U. Frey, L. M. Peticolas, and N. Østgaard (2003b), FAST and IMAGE-FUV observations of a substorm onset, *J. Geophys. Res.*, *108*(A9), 1344, doi:10.1029/2002JA009787.

Newell, P. T., C.-I. Meng, and K. M. Lyons (1996), Suppression of discrete aurorae by sunlight, *Nature*, *381*, 766–767, doi:10.1038/381766a0.

Newell, P. T., T. Sotirelis, and S. Wing (2009), Diffuse, monoenergetic, and broadband aurora: The global precipitation budget, *J. Geophys. Res.*, *114*, A09207, doi:10.1029/2009JA014326.

Østgaard, N., and K. M. Laundal (2012), Auroral asymmetries in the conjugate hemispheres and interhemispheric currents, in *Auroral Phenomenology and Magnetospheric Processes: Earth and Other Planets, Geophys. Monogr. Ser.*, vol. 197, edited by A. Keiling et al., pp. 99–111, AGU, Washington, D. C., doi:10.1029/2011GM001190.

Østgaard, N., S. B. Mende, H. U. Frey, L. A. Frank, and J. B. Sigwarth (2003), Observations of non-conjugate theta aurora, *Geophys. Res. Lett.*, *30*(21), 2125, doi:10.1029/2003GL017914.

Østgaard, N., K. M. Laundal, L. Juusola, A. Åsnes, S. E. Haaland, and J. M. Weygand (2011), Interhemispherical asymmetry of substorm onset locations and the interplanetary magnetic field, *Geophys. Res. Lett.*, *38*, L08104, doi:10.1029/2011GL046767.

Ohtani, S., S. Wing, G. Ueno, and T. Higuchi (2009), Dependence of premidnight field-aligned currents and particle precipitation on solar illumination, *J. Geophys. Res.*, *114*, A12205, doi:10.1029/2009JA014115.

Paschmann, G., S. Haaland, and R. Treumann (2002), *Auroral Plasma Physics*, 100–120 pp., ISSI, Bern.

Press, W. H., S. A. Teukolsky, W. T. Vetterling, and B. P. Flannery (1992), *Numerical Recipes in Fortran 77: The Art of Scientific Computing*, 2nd ed., pp. 614–622, Cambridge Univ. Press, Cambridge, U. K.

Reistad, J. P., N. Østgaard, K. M. Laundal, and K. Oksavik (2013), On the non-conjugacy of nightside aurora and their generator mechanisms, *J. Geophys. Res. Space Physics*, *118*, 3394–3406, doi:10.1002/jgra.50300.

Sato, N., T. Nagaoka, K. Hashimoto, and T. Saemundsson (1998), Conjugacy of isolated auroral arcs and nonconjugate auroral breakups, *J. Geophys. Res.*, *103*(A6), 11,641–11,652, doi:10.1029/98JA00461.

Shue, J.-H., P. T. Newell, K. Liou, and C.-I. Meng (2001), Influence of interplanetary magnetic field on global auroral patterns, *J. Geophys. Res.*, *106*(A4), 5913–926.

Shue, J.-H., P. T. Newell, K. Liou, C. Meng, and S. W. H. Cowley (2002), Interplanetary magnetic field Bx asymmetry effect on auroral brightness, *J. Geophys. Res.*, *107*(A8), SIA 16-1–SIA 16-1, doi:10.1029/2001JA000229.

Siscoe, G. L., W. Lotko, and B. U. Ö. Sonnerup (1991), A high-latitude, low-latitude boundary layer model of the convection current system, *J. Geophys. Res.*, *96*(A3), 3487–3495.

Siscoe, G. L., N. U. Crooker, G. M. Erickson, B. U. O. Sonnerup, K. Siebert, D. R. Weimer, W. W. White, and N. C. Maynard (2000), Global geometry of magnetospheric currents inferred from MHD simulations, in *Magnetospheric Current Systems, Geophys. Monogr. Ser.*, vol. 118, edited by S. Othani et al., pp. 41–52, AGU, Washington, D. C.

Song, Y., and R. Lysak (2001), The physics in the auroral dynamo regions and auroral particle acceleration, *Phys. Chem. Earth Part C*, *26*(1–3), 33–42, doi:10.1016/S1464-1917(00)00087-8.

Stenbaek-Nielsen, H. C., and T. N. Davis (1972), Relative motions of auroral conjugate points during substorms, *J. Geophys. Res.*, *77*(10), 1844–1858.

Stenbaek-Nielsen, H. C., and A. Otto (1997), Conjugate auroras and the interplanetary magnetic field, *J. Geophys. Res.*, *102*(A2), 2223–2232.

Tanaka, T. (2000), Field-aligned-current systems in the numerically simulated magnetosphere, in *Magnetospheric Current Systems, Geophys. Monogr. Ser.*, vol. 118, edited by S. Othani et al., pp. 53–59, AGU, Washington, D. C.